\documentclass[letterpaper]{article}

\usepackage{amsmath}
\usepackage{natbib,alifeconf}  %% The order is important
\usepackage{url,hyperref,cleveref}
\usepackage{booktabs}
\usepackage{amsfonts}
\usepackage{amssymb}
\usepackage{graphicx}

\title{Using Dynamical Systems Theory to Quantify Complexity in Asymptotic Lenia}

\author{
    Ivan Yevenko$^1$, 
    Hiroki Kojima$^{2,3}$, \and
    Chrystopher L. Nehaniv$^1$ \\
    \mbox{}\\
    $^1$University of Waterloo, Canada \\
    $^2$The University of Tokyo, Japan \\
    $^3$Alternative Machine Inc., Japan \\
    iyevenko@uwaterloo.ca \\
} % email of corresponding author

\begin{document}
\maketitle 

\begin{abstract}
    % Abstract length should not exceed 250 words
    Continuous cellular automata (CCAs) have evolved from discrete lookup tables to continuous partial differential equation (PDE) formulations in the search for novel forms of complexity. Despite innovations in qualitative behavior, analytical methods have lagged behind, reinforcing the notion that emergent complexity defies simple explanation. In this paper, we demonstrate that the PDE formulation of Asymptotic Lenia enables rigorous analysis using dynamical systems theory. We apply the concepts of symmetries, attractors, Lyapunov exponents, and fractal dimensions to characterize complex behaviors mathematically. Our contributions include: (1) a mathematical explanation for the four distinct solution classes (solitons, rotators, periodic and chaotic patterns), (2) conditions for the existence of a global attractor with fractal dimension $>4$, (3) identification of Kaplan-Yorke dimension as an effective complexity measure for CCAs, and (4) an efficient open-source implementation for calculating Lyapunov exponents and the covariant Lyapunov vectors for CCAs. We conclude by identifying the minimal set of properties that enable complex behavior in a broader class of CCAs. This framework provides a foundation for understanding and measuring complexity in artificial life systems.

\end{abstract}

% If sharing code / data, anonymize your repository and paste the link here.
% Example of anonymizing sevice for github: https://anonymous.4open.science/
% delete this line if not needed
Data/Code available at: \url{https://github.com/iyevenko/DynamicalCA}

\section{Background}\label{sec:background}

\subsection{Continuous Cellular Automata}

\begin{figure}[ht]
    \centering
    \includegraphics[width=\linewidth]{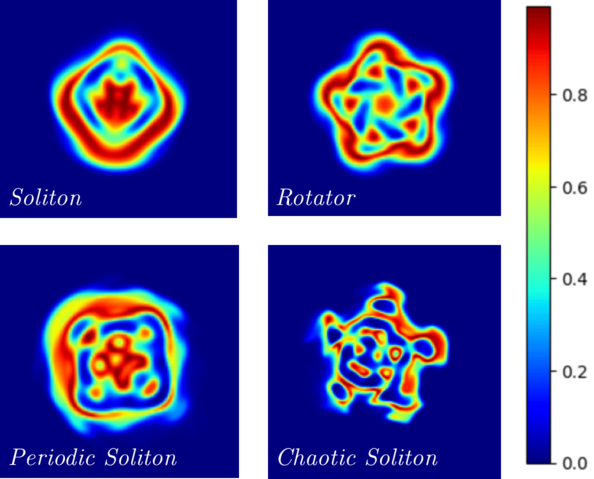}
    \caption{Four classes of dynamic solutions to Asymptotic Lenia, which exist in distinct regions of the PDE's parameter space. The \textit{soliton} is a stable, purely translating solution. The \textit{rotator} is a stable, purely rotating solution. The \textit{periodic soliton} is a translating solution which also oscillates periodically. The \textit{chaotic soliton} exhibits a chaotic combination of translations, rotations, and internal oscillations.}
    \label{fig:lenia_zoo}
\end{figure}

Continuous cellular automata (CCAs) were initially invented to recreate the complex emergent behavior of the program known as Conway's Game of Life \citep{Gardner1970}, with the hope of finding more life-like patterns. In the search for complexity, CAs gradually evolved towards a pure (non-local) partial differential equation (PDE) formulation. ``Larger than life'' \citep{Evans2001} increased the size of the kernel, SmoothLife \citep{Rafler2011} introduced continuous states, Lenia \citep{Chan2019} introduced continuous time and space, and finally Asymptotic Lenia \citep{alenia2021} removed the discontinuous update rule. Despite this total transition from lookup table to PDE, the methods for analyzing these systems have seen very little progress. As a result, the field has largely retained the intuition that emergent complexity has no simple causal explanation.

In this paper, we show that transitioning to a PDE model of CCAs allows us to import well-established mathematical concepts from dynamical systems theory. We first introduce the main concepts: symmetries, attractors, Lyapunov exponents, fractal dimension, and more. Using these concepts, we establish a mathematical explanation for the observed complexity in Asymptotic Lenia. We then measure the relevant quantities to obtain a complete characterization of the complexity of solutions. Finally, we suggest a specific quantity for measuring complexity in CCAs and summarize the small set of key properties which are sufficient for a broader class of CCAs to exhibit complex behavior.

\subsection{Asymptotic Lenia}

Our analysis centers around the Asymptotic Lenia system since it is the only CCA discovered thus far that is defined as a continuous PDE \citep{Davis2022,Kojima2023}. The main classes of dynamic solutions to the PDE are shown in \cref{fig:lenia_zoo}. The PDE is defined by \citet{alenia2021} as follows:

\begin{equation}
    \partial_t A(\mathbf{x},t) = \mathbf{F}(A,K*A) = T(K*A) - A, \label{asymptotic_lenia}
\end{equation}

\noindent where $A \colon \mathbb{R}^2 \times \mathbb{R}^+ \to \mathbb{R}$ is a real-valued 2-D scalar field, $K \colon \mathbb{R}^2 \to \mathbb{R}$ is a radially symmetric kernel, $(*)$ is the 2-D convolution operator, and $T \colon \mathbb{R} \to \mathbb{R}$ is the growth function, applied pointwise. The Gaussian growth function $T(u)=\exp(-\tfrac{1}{2}(u-\mu)^2/\sigma^2)$ is parameterized by $0<\sigma \ll \mu$, and is typically approximated via a polynomial or step function for computational efficiency. The kernel $K(\mathbf{x})=K(\|\mathbf{x}\|_2)$ is composed of a sum of uniformly spaced concentric rings of the form $K_C(r)=(4r(r-1))^4$ with parameters $\{\beta_i\}$ controlling the scaling of the rings, and $R$ controlling the overall spatial scale. Importantly, $K(0)=0$ and $K$ is normalized such that $\int_{\Omega} K(\mathbf{x})\,d\mathbf{x}=1$. We assume that the PDE is solved on a bounded domain $\Omega\subset\mathbb{R}^2$ with periodic boundary conditions.

We also define the spatially-discretized version of Asymptotic Lenia, which is a finite-dimensional (ODE) approximation of the PDE, and is closer to the true numerical implementation of the system. At each discrete point in space $(i,j) \in [0,N-1]^2$, we define a function of time $A[i,j](t)\in\mathbb{R}$. The field's dynamics are given by:

\begin{equation}
\frac{d}{dt} A[i,j](t) = \mathbf{F}[i,j](A) = T\left((K*A)[i,j]\right) - A[i,j]. \label{discrete_asymptotic_lenia}
\end{equation}

\subsection{Dynamical Systems}

To analyze Asymptotic Lenia, we seek a theoretical framework that describes the evolution of fields---functions of space---over time. This is the subject of the theory of infinite-dimensional dynamical systems. In this paper, however, we restrict our analysis to the finite-dimensional setting because it makes our results easier to follow, without loss of generality. A fully rigorous infinite-dimensional treatment of CCAs is beyond the scope of this paper, but we do borrow one important concept: symmetry.

A \textit{symmetry group} of a system of differential equations is a group of transformations that maps solutions of the equations to other solutions \citep{Olver1986}. If the transformation depends on a single continuous parameter (e.g. a shift by $\delta$ in the $x$-direction), the symmetry is called a \textit{continuous symmetry}. Continuous symmetries can be organized into a \textit{Lie group}. An example of a Lie group is the set of rotations and translations on a 2-D plane $SE(2)$. The associated set of infinitesimal transformations is called the \textit{Lie algebra} and it is composed of the \textit{generators} of the group. For example, the generators for translation in the $x$-direction is $\partial_x$ and the generator for rotation is $x\partial_y - y\partial_x$. All 2-D CCAs are invariant under the action of the group $SE(2)$ because the update rule is identical across all of space, and the non-local component is only a function of radial distance. They are also invariant under time translation since the update rule is autonomous, meaning it is independent of time.

Now, we describe the finite-dimensional setting for dynamical systems theory in the language of \citep[e.g.,][]{Temam1997,Hale1988,Robinson2001,Sell2002}. The definitions are mapped to discretized Asymptotic Lenia as follows:

\begin{itemize}
    \item \textit{Phase Space}: $\mathcal{X}=\mathbb{R}^{N^2}$: The set of all possible states of the system. 
    \item \textit{Evolution Law}: $\dot A(t)=\mathbf{F}(A)$: The system of ODEs that describes the evolution of the state over time.
    \item \textit{Semiflow} $\Phi_t(A(0))=A(t),t\ge 0$: The map that takes an initial state $A(0)$ and a time $t$ to the future state $A(t)$. Its existence is guaranteed because $\mathbf{F}$ is Lipschitz continuous and $|A|$ can never blow up in finite time \citep{Robinson2001}\footnote{The rigorous proof is omitted. The Lipschitz property follows from that fact that $K*A$ is bounded and $T$ is Lipschitz. $|A|$ is bounded because it can at most grow/shrink exponentially.}.
\end{itemize}

With the above definitions in place, we can discuss the following properties of the dynamical system.

\textit{Dissipativity}: A semiflow $\Phi_t$ over $\mathbb{R}^n$ is dissipative if there exists a bounded set $\mathcal{B} \subset \mathbb{R}^n$ such that every initial condition $A(0)\in\mathbb{R}^n$ is absorbed into $\mathcal{B}$ in a finite time $t^*$.

\textit{Global Attractor}: If a semiflow $\Phi_t$ over $\mathbb{R}^n$ is dissipative, then there exists a global attractor $\mathcal{A}$ which is the minimal set onto which all trajectories converge as $t\to\infty$. The attractor entirely contains the long-term behavior of the system after transients have died away. A global attractor also guarantees at least one invariant measure on the attractor, letting us take physically meaningful integrals over the attractor. 

\textit{Fractal Dimension}: Dissipative chaotic dynamical systems tend to settle to attractors which integrate to zero phase space volume but cannot be represented as an integer-dimensional manifold. These are called fractal attractors, and they are characterized by their fractal dimension. Fractal dimension approximately represents the number of degrees of freedom in a given set, and there are several relevant definitions for fractal dimension. We focus on specific cases of the generalized dimensions introduced by \citet{Grassberger1983}. They are defined as:

\[
D_q = \frac{1}{1-q} \lim_{\varepsilon\to 0} \frac{\log (\sum_{i} p_i^q)}{\log (1/\varepsilon)},
\]

\noindent where $p_i$ is the total mass in the $i$-th box in a uniform partition of phase space with some resolution $\varepsilon$. $D_0$ is the \textit{box-counting dimension}, $D_1$ is the \textit{information dimension}, and $D_2$ is the \textit{correlation dimension}. These only differ when the distribution of density on the attractor is non-uniform.

\textit{Lyapunov Exponents}: The Lyapunov exponents (LEs) $\{\lambda_i\}_{i=1}^n$ of a trajectory $A(t)$ are defined as the average exponential growth rates of perturbations to $A(0)$. If there exists at least one positive Lyapunov exponent, then we call the system chaotic because nearby trajectories exponentially diverge. There exist several definitions for LEs (see \citet[Sec.~6.3.1]{Kuznetsov2020} for a comprehensive overview), but the one that applies best to our setting is: 
\[
\lambda_i(A(0))=\limsup_{t\to \infty} \frac{1}{t} \log \sigma_i(A(0),t),
\]
\noindent where $\sigma_i$ is the $i$-th largest singular value of the Jacobian $\nabla_A \Phi_t(A)$. These exponents can be approximated numerically using the method presented by \citet{Benettin1980}. Furthermore, the covariant Lyapunov vectors, which encode the associated perturbation directions for each exponent, can be calculated easily with a short backwards pass after applying Benettin's method \citep{Ginelli2007}.

\textit{Kaplan-Yorke Dimension}: The Kaplan-Yorke dimension $D_{KY}$ is a heuristic upper bound on the information dimension of the attractor. Experimentally, it has been shown to closely estimate $D_1$, except in rare edge cases (see discussion in \citep{Groger2013}). It is defined as
\[
D_{KY}=j + \frac{\sum_{i=1}^j \lambda_i}{|\lambda_{j+1}|},
\]
\noindent where $j$ is the largest index such that the cumulative sum of the largest $j$ Lyapunov exponents is non-negative.

\section{Mathematical Results}
\subsection{Existence of Symmetry Solutions}

As explained in the Background section, \eqref{asymptotic_lenia} is invariant under the action of the group $G=SE(2)\times\mathbb{R}$. The corresponding Lie algebra is:
\[
\mathfrak g=\operatorname{span}_{\mathbb R}\{\partial_x, \partial_y, x\partial_y-y\partial_x, \partial_t\}
\]

Using these symmetries, we can derive conditions under which specific forms of solutions exist. We refer to these as \textit{symmetry solutions}, and we can express them with the ansatz $A(\mathbf{x},t)=e^{t \gamma}A_0$, where $\gamma \in \mathfrak{g}$ and $A_0=A(\mathbf{x},0)$. This ansatz becomes a solution when the following equation is satisfied:

\begin{align}
    \partial_t\left[e^{t \gamma} A_0\right] &= \mathbf{F}\left(e^{t \gamma} A_0, K * (e^{t \gamma} A_0)\right) \nonumber \\
    \gamma\,e^{t \gamma} A_0                &= \mathbf{F}\left(e^{t \gamma} A_0, e^{t \gamma}(K*A_0)\right) \nonumber \\
    e^{t \gamma}\,\gamma A_0                &= e^{t \gamma}\,\mathbf{F}\left(A_0, K*A_0\right) \nonumber \\
    \gamma A_0                              &= \mathbf{F}\left(A_0, K*A_0\right) \label{stationary_cond}
\end{align}

From the second to third line, we use the $G$-equivariance property of $\mathbf{F}$. We also use the fact that $e^{t\gamma}$ commutes with $\gamma$ and $K$ and has an inverse operator $e^{-t\gamma}$. These facts are not trivial and are a direct consequence of the specific form of the PDE. 

A key result of \eqref{stationary_cond} is that no evolution is required to verify a symmetry solution. For example, to find a translating solution $A(\mathbf{x},t)=A(\mathbf{x}-\mathbf{v}t,0)=e^{t \gamma}A_0,\,\mathbf{v} \in \mathbb{R}^2$, we need only find an initial field $A_0$ such that \eqref{stationary_cond} is satisfied with $\gamma=-\mathbf{v}\cdot\begin{bmatrix}\partial_x\,\, \partial_y\end{bmatrix}^\top$. A similar argument applies to rotating solutions and steady state solutions. Note that since $\mathbf{F}(\mathbf{0}, K*\mathbf{0})=\mathbf{0}$, a trivial solution to \eqref{stationary_cond} always exists. 

There is also an important connection between the Lie algebra and the Lyapunov exponents of the system. Theorem 2.15 of \citet{Aston1994} showed that for $G$-equivariant flows, the number of Lyapunov exponents equal to zero is $\dim\{G\} - \dim\{g \in G\,|\,g\!\cdot\!A=A\}$. In our case, the number of zero exponents equals 4 minus the number of independent continuous symmetries that leave the solution unchanged. For example, a purely translating solution will have at most $4-1=3$ zero exponents because one dimension of the symmetry group (corresponding to translation in the direction of motion combined with an appropriate time shift) acts trivially on the solution.

\subsection{Dissipativity and Existence of a Global Attractor}

In this section, we prove a fundamental property of Asymptotic Lenia: all trajectories remain bounded within a finite region of phase space and eventually converge to a common attractor. This property, known as dissipativity, is crucial for understanding the long-term behavior of the system and ensures that the complexity we observe is constrained within a well-defined mathematical structure.

Using the formal definition from \cite{Robinson2001}, we can show that discretized Asymptotic Lenia is a dissipative semiflow.  For some $\varepsilon > 0$, consider the compact set $\mathcal{B}_\varepsilon=[0-\varepsilon,1+\varepsilon]^{N^2} \subset \mathcal{X}$. Let $m(t)=\min_{i,j}A[i,j](t)$, $M(t)=\max_{i,j}A[i,j](t)$. If we choose the initial condition $A[i,j](0)\in \mathcal{X}$ and assume $0\le T(\cdot)\le 1$, we find:

\begin{align*}
    \frac{dM}{dt} &= T((K*A)[i_M,j_M]) - M \le 1 - M \\[4pt]
                  &\implies M(t) \le 1 + \bigl(M(0)-1\bigr)e^{-t} \\[8pt]
    \frac{dm}{dt} &= T((K*A)[i_m,j_m]) - m \ge -m \\[4pt]
                  &\implies m(t) \ge m(0)\,e^{-t}
\end{align*}

\noindent where $(i_M,j_M)$ and $(i_m,j_m)$ are the indices of the maximum and minimum values of $A[i,j](t)$ respectively. Therefore, after a finite time

\[
t^*= \max\left\{\ln\left|\frac{M(0)-1}{\varepsilon}\right|, \ln\left|\frac{m(0)}{\varepsilon}\right|\right\},
\]
we must have $A[i,j](t)\in \mathcal{B}_\varepsilon,\, \forall\, t \ge t^*$. We conclude that there exists a global attractor $\mathcal{A}$. A very similar argument can be used to show that the full PDE is also dissipative, but this is not necessary for our purposes. We can now inquire about the kinds of behavior we expect to see on the attractor. 

\subsection{Helmholtz Decomposition and Behavior on the Attractor}

In this section, we analyze the mathematical structure of the Asymptotic Lenia dynamics to understand why it can produce complex behaviors such as chaos and periodic oscillations. We show that the system can be decomposed into two fundamental components: a divergence-free (rotational) part and a gradient part. This decomposition, known as the Helmholtz-Hodge decomposition, reveals why Asymptotic Lenia can exhibit rich dynamics rather than simply settling into static patterns. We start by calculating the Jacobian of $\mathbf{F}$ from \eqref{discrete_asymptotic_lenia} as follows:

\begin{align*}
    \frac{\partial \mathbf{F}[i,j](A)}{\partial A[k,l]} &= \frac{\partial}{\partial A[k,l]} \left[T((K*A)[i,j]) - A[i,j]\right]\\
    &= T'((K*A)[i,j])\, \frac{\partial (K*A)[i,j]}{\partial A[k,l]} -  \frac{\partial A[i,j]}{\partial A[k,l]}\\
    &= T'((K*A)[i,j])\, K[i-k,j-l] - \delta_{ik}\delta_{jl}
\end{align*}

Imposing $K[0,0]=0$, the divergence of the first term becomes zero so the divergence comes entirely from summing the Kronecker delta terms, giving $\nabla_A \cdot \mathbf{F} = -N^2$. This means that we can write $\mathbf{F}$ as the sum of a divergence-free term and a term which is the gradient of a scalar potential:
\[
\mathbf{F}[i,j](A) = \underbrace{T\left((K*A)[i,j]\right)}_{\text{divergence-free}} + \underbrace{\frac{\partial}{\partial A[i,j]} \left[-\frac{1}{2} \sum_{k,l} A^2[k,l]\right]}_{\text{gradient}}
\]

Such a decomposition is known as a Helmholtz-Hodge decomposition \citep{hhd2013}. We do not comment on the uniqueness of the decomposition due to the possibility for harmonic components, but we can infer that $\mathbf{F}$ cannot be the gradient of a scalar function. This is because for any arbitrary pair of indices $(i,j)\ne (k,l)$, the entries of the Jacobian are not necessarily equal due to nonlinearity of $T$.

\begin{figure*}[ht!]
    \centering
    \includegraphics[width=\linewidth]{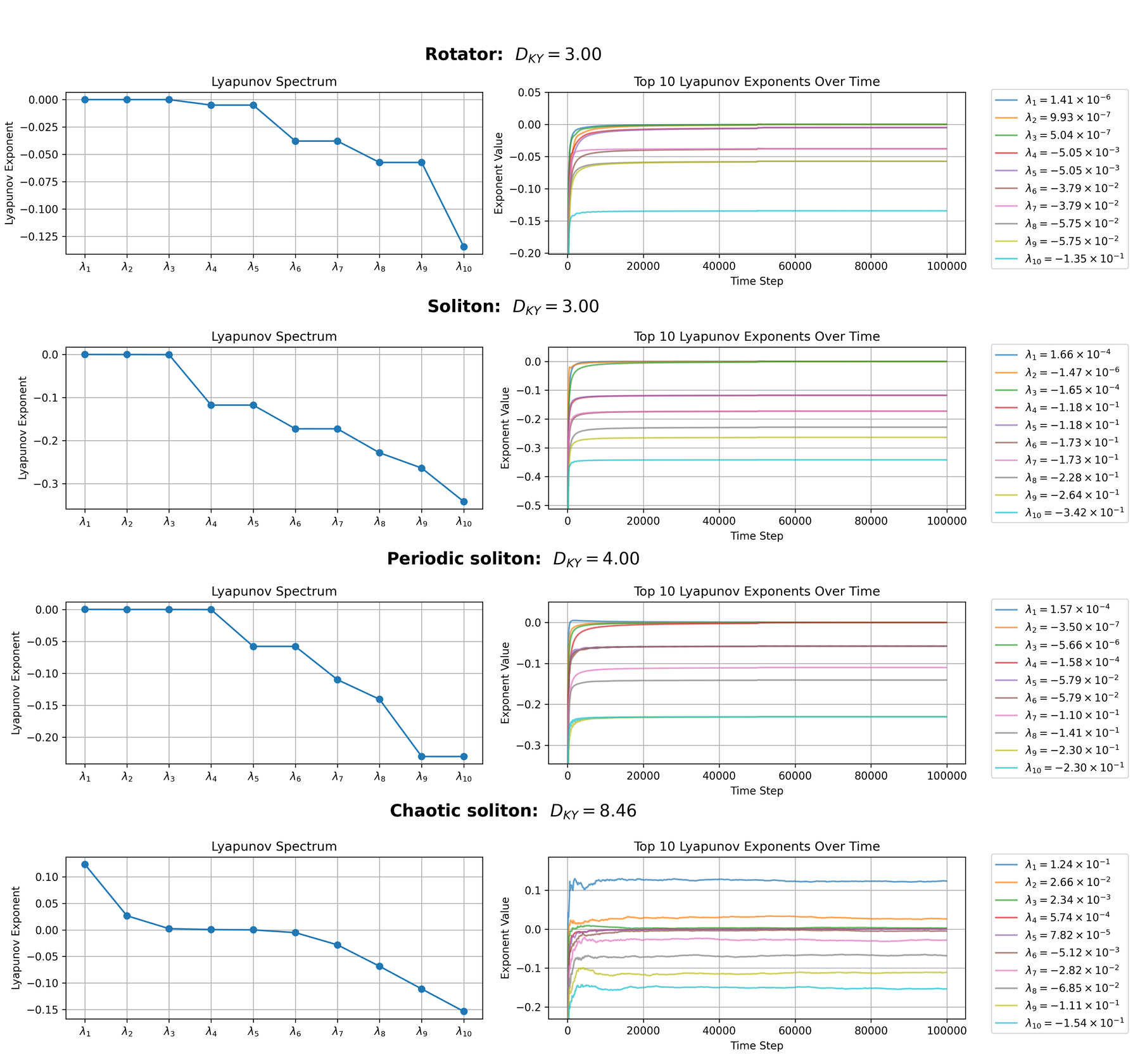}
    \caption{Lyapunov spectra of the four classes of solutions shown in \cref{fig:lenia_zoo}. On the left, we plot the values of the top 10 Lyapunov exponents for each solution class. On the right, we plot the values of the exponents over time to show convergence. The Kaplan-Yorke dimension is calculated based on the values of these exponents and shown in the titles.}
    \label{fig:lyap_spectrum}
\end{figure*}

We stress that this result is crucial for the existence of chaotic or periodic dynamics on attractors. While trajectories of a pure gradient flow are not guaranteed to reach a steady state, they are guaranteed to enter and remain in an arbitrarily small neighborhood of the set of steady states \citep{Hale1988,Robinson2001}. The classic neighbor-sum + non-linearity rule turns out to be exactly the right kind of term to allow for complex behavior.

\begin{figure*}[ht!]
    \centering
    \includegraphics[width=\linewidth]{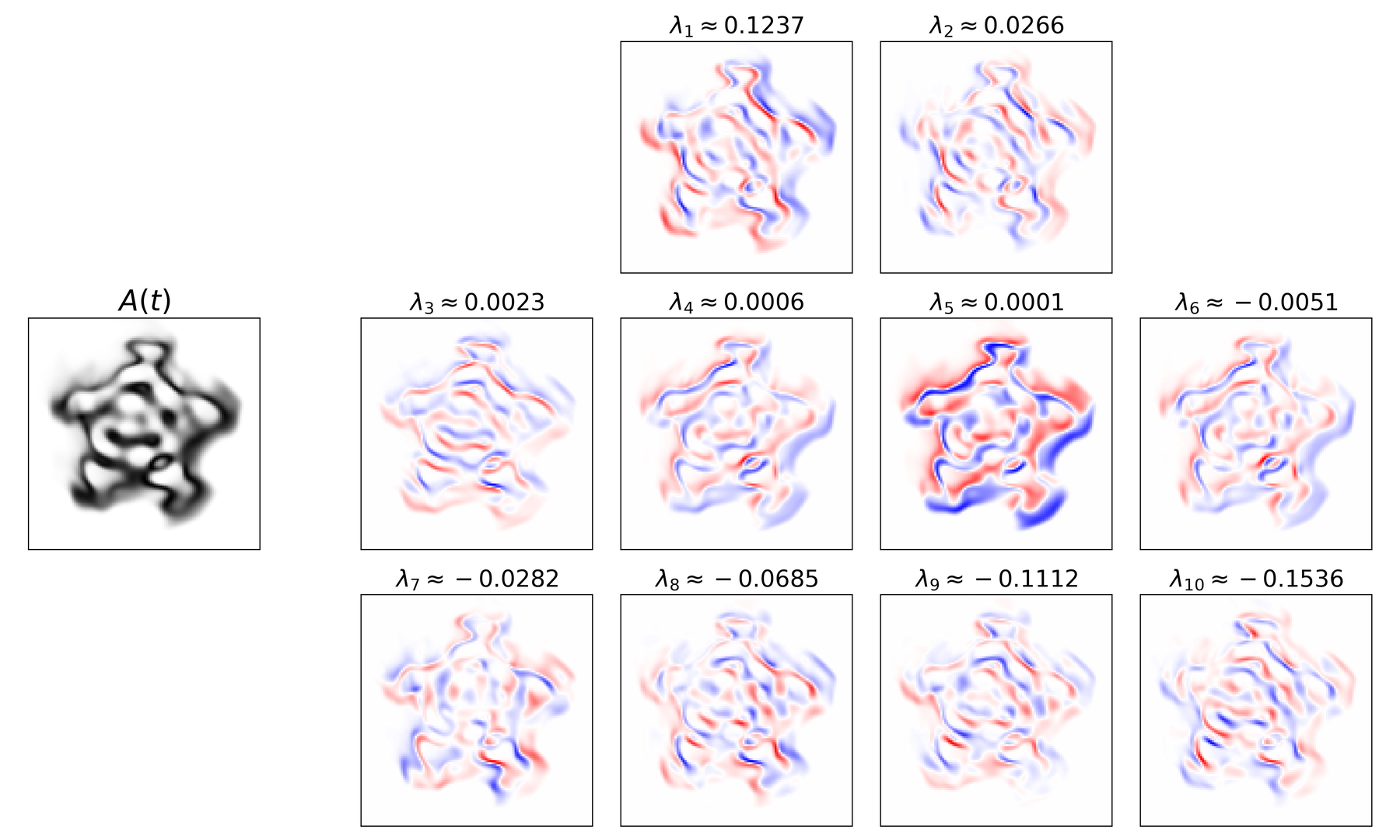}
    \caption{Covariant Lyapunov vectors of the chaotic soliton. On the left in black and white is the value of the field after evolving for 90,000 steps at $\Delta t=0.1$. The top 10 CLVs are visualized on the right, where blue and red correspond to negative and positive values respectively.}
    \label{fig:clv}
\end{figure*}

\section{Experimental Results}

Having established the theoretical framework for analyzing Asymptotic Lenia, we now empirically measure key quantities to characterize the complexity of different solution classes. Our theoretical analysis predicted that chaotic solutions should exhibit a global attractor with fractal dimension between 4 and $N^2$, with the exact dimension reflecting the system's complexity. Here, we verify these predictions by calculating LEs and the corresponding Kaplan-Yorke dimension for representative solutions from each class identified in \cref{fig:lenia_zoo}.

LEs are calculated for the spatially-discretized flow using the method first proposed by \citet{Benettin1980}. In particular, we integrate a set of $k$ orthogonal vectors $\mathbf{v}_i$ along the linearized flow $\dot{\mathbf{v}}_i=\nabla_A \mathbf{F}\, \mathbf{v}_i$ and repeatedly apply QR decomposition to ensure orthogonality. \citet{Benettin1980} showed that the top $k$ exponents equal the time average of the logarithmic growth of the tangent vectors' norms between successive applications of QR decomposition. These top $k$ LEs can then be plugged into the Kaplan-Yorke formula to obtain an upper bound on the information dimension of the attractor.

The Lyapunov spectra for four distinct solution classes are shown in \cref{fig:lyap_spectrum}. We calculate these with a spatial discretization of $N=256$ and a time step of $\Delta t=0.1$. After an initial transient period, we run a forward pass for 100,000 steps and then a backwards pass for 10,000 steps to calculate the LEs and CLVs. Since the tangent vectors $\mathbf{v}_i$ are initialized randomly and take time to converge, we only average over the last 50,000 steps to obtain the finite-time LEs.

As predicted, the purely rotating and translating solutions have 3 zero LEs, while the chaotic and periodic solutions have 4 zero LEs. Intuitively, in all but the chaotic case, the top LEs are all zero, making $D_{KY}$ exactly equal to the number of zero LEs. When the top LEs are all close to zero, the $D_{KY}$ formula is sensitive to tiny differences in the values, so in practice we round all $|\lambda_i|<0.001$ to zero. For the chaotic soliton, we found that it had a Kaplan-Yorke dimension of approximately 8.46, with two significant positive LEs.

The Lyapunov spectra help explain the remarkable robustness to perturbation we observe in practice. Since all but a few of the $N^2$ LEs are negative (with a mean of -1), the system exponentially decays most modes of a random perturbation. In \cref{fig:clv}, we visualize the unstable, neutral, and stable modes by calculating the covariant Lyapunov vectors (CLVs) for the chaotic soliton. It is important to note that the CLVs are not necessarily orthogonal, and in this case they certainly are not. We also observe that within the set of neutral CLVs, $\lambda_4$ and $\lambda_6$ are approximately identical. This is because they theoretically correspond to the same exponent so the CLVs span a degenerate subspace that has no unique decomposition \citep{Ginelli2013}.

\section{Discussion}

In the previous section, we verified that the number of zero LEs matched the theoretically predicted quantity and that chaotic patterns indeed exhibit at least one positive LE and $D_{KY}>4$. The calculation of the LEs was relatively computationally inexpensive and the results were easy to interpret. Compared to other methods of estimating fractal dimension, this ``Lyapunov analysis'' is uniquely effective for reasons we will discuss next.

\subsection{Kaplan-Yorke Dimension as a Measure of Complexity}
Both the box-counting dimension and correlation dimension are impractical measures of fractal dimension for finely discretized PDEs. Box-counting dimension requires constructing histograms with $\mathcal{O}(\varepsilon^{-N^2})$ bins, where $\varepsilon$ is the bin size. Correlation dimension requires calculating the pairwise distance between all sampled fields $A[i,j](n\Delta t)$ on a trajectory, which is expensive but tractable for moderate $N$ and trajectory lengths. However, convergence is slow because the method relies on sampling many points from the tail of the power-law distance distribution. Experimentally, we also found that log-log plots of the pairwise distance distribution did not have a uniquely identifiable linear region, making the correlation dimension difficult to interpret unambiguously. On the other hand, the Kaplan-Yorke dimension requires just a constant number of Jacobian-vector product operations and QR decompositions to be carried out at each timestep of the state evolution. Even for chaotic attractors, we found in our experiments that convergence to within 1\% of the long-time estimate of $D_{KY}$ required only $\sim 10^4$ timesteps at $\Delta t=0.1$. 

Additionally, information dimension (upper bounded by $D_{KY}$) offers the most intuitive interpretation of fractal dimension. It tells us how many bits of information are needed to describe any state on the attractor for a given resolution. Equivalently, we can think of information dimension as answering \textit{how many real numbers are needed to encode the state} as opposed to how many binary numbers---i.e. entropy. For all of these reasons, we conclude that the Kaplan-Yorke dimension is the most practical and informative measure of complexity for Continuous Cellular Automata. 

We suggest that future works on Continuous and Neural CAs use $D_{KY}$ to quantify the complexity of patterns they discover. Next, we discuss how we can systematically search for higher complexity in CCAs.

\subsection{Generalizing CCAs}
We begin by summarizing the mathematical predictions, and the minimal assumptions from which we derived them. This is shown in \cref{tab:assumptions} below.

\begin{table}[h]
\centering
\begin{tabular}{p{0.4\linewidth}p{0.5\linewidth}}
\toprule
\textbf{Assumption} & \textbf{Resulting Prediction} \\
\midrule
$\partial_t A =T(K*A)-A$,\newline $K$ radially symmetric. & Possibility of symmetry solutions and up to four $\lambda_i=0$. \\
\midrule
$0 \le T(\cdot) \le 1$,\newline $T$ Lipschitz continuous & Existence of global attractor. \\ 
\midrule
$K[0,0] = 0$ & $\dim \mathcal{A} < N^2$. Not a gradient flow, possibility of dynamic attractors. \\
\midrule
\bottomrule
\end{tabular}
\caption{Summary of key assumptions and their mathematical implications for Asymptotic Lenia dynamics.}
\label{tab:assumptions}
\end{table}

The listed assumptions alone do not guarantee the existence of a chaotic attractor. However, we know that at least one choice of $T$ and $K$ satisfying the assumptions does produce a chaotic attractor because Asymptotic Lenia is a known example. We also know that Asymptotic Lenia does not span the entire set of such CCAs. The parameterization of $T$ and $K$ via the parameter set $\{\mu, \sigma\}\cup \{\beta_i\}$ is just one low-dimensional parameterization of the space of functions satisfying the listed assumptions. 

We could stop there, and suggest that there are many other forms of $T$ and $K$ which exhibit similar kinds of chaotic behavior. However, we aim to find a broader class of CCAs which might exhibit new types of complex behavior like replication, reproduction, or goal-directedness. We suggest that the following general formulation of CCAs be explored in future work:

\begin{equation}
    \partial_t A(\mathbf{x},t) = f(K*A) - \nabla_A G(A),
\end{equation}

\noindent where $K$ is bounded, radially symmetric, and zero at the origin, and $f$ is a Lipschitz continuous nonlinear function such that $0 \le f(\cdot) \le 1$. If we require $G$ be a convex function with no minimizers outside $[0,1]^{N^2}$ and $\nabla_A G(\mathbf{1}) \ge \mathbf{1}$, we can guarantee both dissipativity and nonpositive divergence. This follows from a similar argument to the one made in the mathematical results section, but we omit the proof. In this formulation, the divergence is still negative since it is the Laplacian of a convex function, but it will also be nonuniform over the phase space. We suspect that nonuniform divergence may be enough for highly multifractal attractors to emerge, leading to novel forms of complexity.

\section{Conclusion}

We have shown that the mathematical framework of dynamical systems theory can be rigorously applied to Asymptotic Lenia, and that it provides a powerful set of tools for understanding complexity in a broader class of CCAs. We introduced Lyapunov exponents and fractal dimension as fundamental concepts and provided a simple open-source implementation of the algorithm used to calculate them. This allowed us to experimentally verify that just a few key properties of the system are sufficient to enable chaotic behavior. Using these properties, we proposed a generalization of CCAs which we believe may contain new forms of complexity which cannot be characterized by a single Lyapunov spectrum.

For future work, we suggest that the full infinite-dimensional dynamical systems framework be applied to our general CCA formulation. In particular, we are very interested in an extended theoretical framework which could explain and predict phenomena like replication, reproduction, and goal-directedness. We also strongly encourage future works in continuous-time artificial life simulations like neural cellular automata to quantify the complexity of their systems using the Lyapunov analysis presented here.

\section*{Acknowledgments}Support of the Natural Sciences and Engineering Research Council of Canada (NSERC), funding reference number RGPIN-2019-04669 is gratefully acknowledged.
Cette recherche a \'et\'e financ\'ee par le Conseil de recherches en sciences naturelles et en g\'enie du Canada (CRSNG), num\'ero de r\'ef\'erence RGPIN-2019-04669. In addition the work of the first author was supported by a University of Waterloo Undergraduate Research Assistantship President's Award.

\footnotesize
\bibliographystyle{apalike}
\bibliography{main}

\end{document}